\documentclass[aip,amsmath,amssymb,reprint,]{revtex4-1}

\usepackage{graphicx}
\usepackage{dcolumn}
\usepackage{bm}

\usepackage[utf8]{inputenc}
\usepackage[T1]{fontenc}
\usepackage{mathptmx}

\begin{document}

\preprint{AIP/123-QED}

\title[Interaction between coaxial dielectric disks enhances the Q-factor ]
{Interaction between coaxial dielectric disks  enhances the Q factor}
\author{K.N. Pichugin}
\affiliation{Kirensky Institute of Physics, Federal Research Center
KSC SB RAS, 660036 Krasnoyarsk, Russia}

%
\author{A. F. Sadreev}%
 \email{almas@tnp.krasn.ru.}
\affiliation{Kirensky Institute of Physics, Federal Research Center
KSC SB RAS, 660036 Krasnoyarsk, Russia}
\date{\today}

\begin{abstract}
We study the behavior of resonant modes under variation of the
distance between two coaxial dielectric disks and show an avoided
crossing of resonances because of interaction between the disks.
Owing to coaxial arrange of disks the consideration is separated
by the azimuthal index $m=0, 1, 2, \ldots$. In the present paper
we consider the case $m=0$. For a long enough distance the
resonant modes can be classified as symmetric and antisymmetric
hybridizations of the resonant modes of the isolated
disk. With decreasing of the distance the interaction becomes stronger that gives rise
to avoided crossing of different resonances of the isolated disk.
That in turn enhances the $Q$ factor of two disks
by one order in magnitude compared to the $Q$ factor of isolated disk.
\end{abstract}

\maketitle

%

\section{Introduction}

Optical microcavities and various other sorts of resonators have
been widely employed to tightly localize electromagnetic field in
small volumes for a long durations due to high Q factors, which
plays an indispensable role in lasing, sensing, filtering and many
other applications in both the linear and nonlinear regimes. In
the cavity the local photon density of states scales
proportionally to the quality factor to volume of the cavity
ratio.  In general,there is a compromise between high $Q$ factors
and small mode volumes due to the fact that larger resonators are
required to increase round-trip travel time for $Q$ factor
enhancement, as is the case for whispering gallery modes
\cite{Braginsky1989,Cao2015}.

It is rather challenging for optical resonators to support
resonances of simultaneous subwavelength mode volumes and high
$Q$ factors. The traditional way for increasing of the $Q$ factor of
optical cavities is a suppression of leakage of resonance mode
into the radiation continua.   That is achieved usually by
decreasing the coupling of the resonant mode with the continua.
However, microcavities and resonators based on their reflection
from the boundaries demonstrate substantially low values of the
$Q$ factor by virtue of weakness of the dielectric contrast of
optical materials. The conventional ways to realize high-$Q$
resonators are the use of metals, photonic bandgap structures, or
whispering-gallery-mode resonators. All of these approaches lead
to reduced device efficiencies because of complex designs,
inevitable metallic losses, or large cavity sizes. On the
contrary,   all-dielectric   subwavelength nanoparticles  have
 recently   been   suggested   as   an   important   pathway   to
enhance capabilities of traditional nanoscale resonators by exploiting the multipolar Mie resonances being  limited
only by radiation losses \cite{Kuznetsov2016,Koshelev2018}.

The decisive breakthrough came with the paper by Friedrich and
Wintgen \cite{Friedrich1985} which put forward the idea of
destructive interference of two neighboring resonant modes leaking
into the continuum. Based on a simple generic two-level model they
formulated the condition  for the bound state in the continuum
(BIC) as the state with zero resonant width for crossing of
eigenlevels of the cavity. This principle was later explored in
open plane wave resonator where the BIC occurs in the vicinity of
degeneracy of resonance frequencies \cite{SBR}.

However, these BICs exist provided that they embedded into a
single continuum of propagating modes of a directional waveguide.
In photonics the optical BICs embedded into the radiation
continuum can be realized by two ways. The first way is realized
in an optical cavity coupled with the continuum of 2d photonic
crystal (PhC) waveguide \cite{BS2008} that is an optical variant
of microwave system \cite{SBR}. More perspective way is to achieve
the BICs in periodic PhC systems or arrays of dielectric particles
in which resonant modes leak into a restricted number of
diffraction continua
\cite{Shipman2005,Marinica,Hsu2013,PRA2014,PRA96}. Although the
exact BICs can exist only in infinite periodical arrays
\cite{Colton,Silveirinha14}, the finite arrays demonstrate
resonant modes with the very high $Q$ factor which grows
quadratically with the number of particles \cite{Balyzin2018}
(quasi-BICs).

Another attractive way to achieve quasi-BICs (super cavity modes)
is to use individual subwavelength high-index dielectric
resonators, which exhibit also high-$Q$ factors
\cite{Rybin2017,Koshelev2018,WeijinChen2018,Bogdanov2019}. Such
super cavity modes originate from avoided crossing of the nearest
resonant modes, specifically the TE (Mie-type) resonant mode and
the TM (Fabry-P\'{e}rot) resonant mode under variation of the
aspect ratio of the dielectric disk. As the result they report a
significant enhancement of the $Q$ factor. It is worthy also to
notice  the idea of formation of long-lived, scar like modes near
avoided resonance crossings in optical deformed microcavities
\cite{Wiersig2006}. The dramatic  $Q$ factor enhancement was
predicted by Boriskina \cite{Boriskina2006,Boriskina2007} for
avoided crossing of highly excited whispering gallery modes in
symmetrical photonic molecules of dielectric cylinders.

In the present paper we consider a similar way to enhance the
$Q$ factor by variation of the distance between two identical
coaxial dielectric disks. As different from consideration in papers
\cite{Wiersig2006,Boriskina2006,Boriskina2007,Unter2008,Benyoucef2011}
we consider the avoided crossing of
low excited resonant modes (monopole and dipole) with variation of
the distance between two dielectric disks. When the disks are
separated by the enough long distance we have pairs of
degenerate resonant modes such as monopole, dipole, {\it etc}.
With the decrease of the distance the resonant modes interfere
given rise to avoided crossing. We show that this effect is
complimenting by a spiral behavior of the resonant eigenvalues
when the interaction between the disks is weak. With a further
decrease of the distance the interaction is increasing to give
rise to strong repulsion of the resonances. For this phenomenon
one of the resonant eigenfrequencies can approach to the real axis
distinguishing the high $Q$ factor compared to the $Q$ factor of
isolated disk.

\section{Avoided crossing in system of two coaxial disks}
\begin{figure}
\includegraphics[width=6cm,clip=]{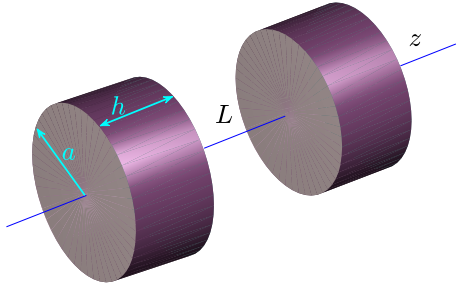}
\caption{Two coaxial dielectric disks separated by distance $L$
referred between the centers of disks.} \label{fig1}
\end{figure}

There are two limiting cases of the system of coaxial dielectric
disks, the infinite periodic array of disks and the isolated disk.
The former supports rich variety of resonant modes with zero resonant widths,
BICs: symmetry protected BICs
in the $\Gamma$-point, accidental BICs with nonzero Bloch vector,
and hybrid BICs with nonzero orbital angular momentum
\cite{Bulgakov2017}. The symmetry protected BICs were
experimentally observed in the system of ceramic disks in the THz
range \cite{Balyzin2018}. The case of the isolated disk was
considered in papers \cite{Rybin2017,Koshelev2018} which have
shown considerable enhancement of the $Q$ factor due to an avoided
crossing of two resonant modes. In that papers the avoided
crossing was achieved by a variation of aspect ratio of the
disk that technologically is not simple except. In the present section we
consider two identical coaxial disks as sketched in Fig. \ref{fig1}
each of them haven the aspect ratio not obligatory tuned to the
optimal Q-factor as in \cite{Rybin2017,Bogdanov2019}. The coaxial
disks have the advantage that all resonant modes are classified by
the azimuthal number $m=0, 1, 2, \ldots$ because of the axial
symmetry. Therefore one can consider subspaces with definite $m$
separately. In the present paper we follow the case $m=0$ for
which the solutions are separated by polarization with $H_z=0$ (E
modes) and $E_z=0$ (H modes). In what follows we consider the
H-modes.

\begin{figure}
\includegraphics[width=9cm,clip=]{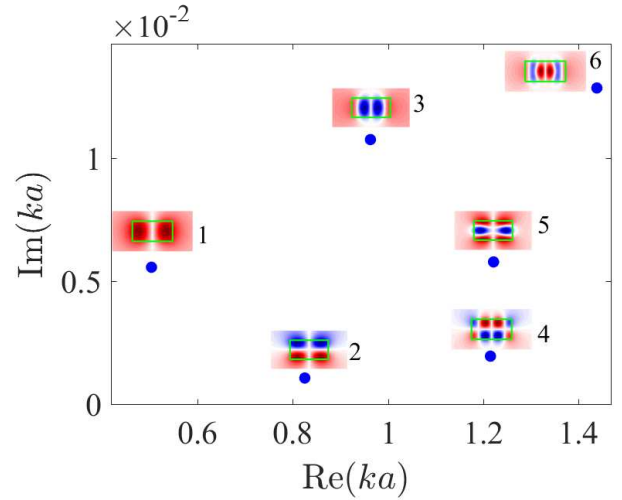}
\caption{The resonant eigenfrequencies (close circles) and
corresponding resonant modes (the component $E_{\phi}$ of
dielectric disk with the height $h=a$ and permittivity
$\epsilon=40$.}
 \label{fig2}
\end{figure}

In general the resonant modes and their eigenfrequencies are given by
solving the time-harmonic source-free Maxwell's equations \cite{Joan,Lalanne2018}
\begin{equation}\label{ME}
    \left(%
\begin{array}{cc}
  0 & \frac{i}{\epsilon}\nabla\times \\
  -i\nabla\times & 0\\
\end{array}%
\right)
\left(%
\begin{array}{c}
  \mathbf{E}_n \\
  \mathbf{H}_n \\
\end{array}%
\right)=k_n\left(%
\begin{array}{c}
  \mathbf{E}_n \\
  \mathbf{H}_n \\
\end{array}%
\right)
\end{equation}
where $\mathbf{E}_n$ and $\mathbf{H}_n$ are the EM field
components defined in Ref. \cite{Lalanne2018} as quasinormal modes
which are also known as resonant states
\cite{More1973,Muljarov2010} or leaky modes \cite{Snyder1984}. It
is important that they can be normalized and the orthogonality
relation can be fulfilled by the use of perfectly matched layers
(PMLs) \cite{Lalanne2018}. With the exception very restricted
number of symmetrical particles (cylinders, spheres) Eq.
(\ref{ME}) can be solved only numerically. Irrespective to the choice of dielectric particle
the eigenfrequencies are complex $k_na=\omega_n+i\gamma_n$ where $a$ is
the disk radius. In what follows the light velocity is taken unit.
 Fig. \ref{fig2} shows resonant frequencies of the single isolated disk
complimented by insets with the resonant modes (only the component
$E_{\phi}$ is shown). There are modes with nodal surfaces crossing
the z-axis  and the modes with nodal surfaces crossing the plane
$z=0$. They correspond to the Fabry-Perot resonant modes and the
radial Mie modes by the terminology introduced in paper \cite{Rybin2017}.

\begin{figure}
\includegraphics[width=9cm,clip=]{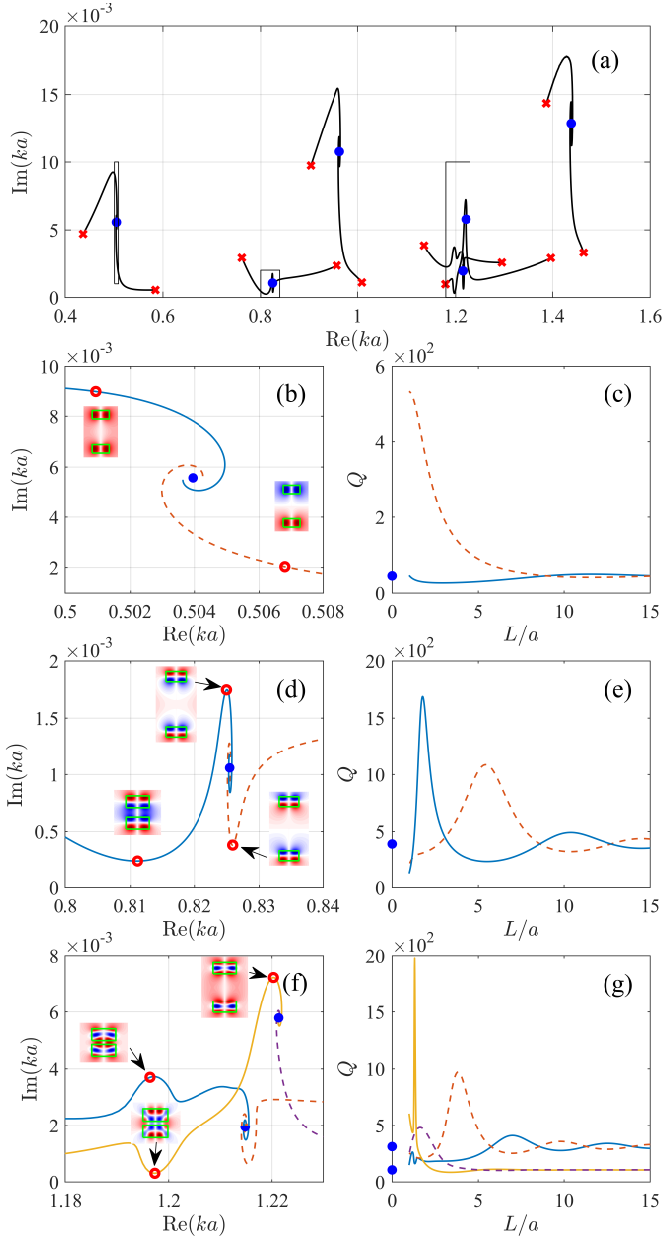}
\caption{(a) Behavior of resonant eigenfrequencies under variation
of the distance between the disks $L$ with the same parameters as in Fig. \ref{fig2}.
(b) and (d) zoomed
areas highlighted in (a) with symmetric (solid lines) and antisymmetric (dash lines)
hybridization (\ref{s,a}) of resonant modes of the isolated disk.
(c) and (e) show behavior of the $Q$ factor vs the distance for
corresponding insets at the left. Closed circles mark the eigenfrequencies of isolated disks and
respectively the $Q$ factors while crosses mark the limiting case $L=h$ when two disks stick together.}
\label{fig3}
\end{figure}

Fig. \ref{fig3} shows the solutions of Eq. (\ref{ME}) for the case
of two coaxial dielectric disks as dependent on the distance $L$
between the disks.
The necessity to use PMLs restricts the distance
between the disks which is to be considerably less than the
distance between the PMLs in the z-direction.
In spite of an illusive complexity in Fig. \ref{fig3} the zoomed
pictures reveal remarkably simple behavior of resonant frequencies
in the form of a spiral convergence of avoided eigenfrequencies to
the resonant frequencies of the isolated disks marked by closed
circles as Fig. \ref{fig3} demonstrates at zoomed plots. However
when the disks approach close enough to each other the spiralling
behavior is substituted by strong repulsion of resonant
frequencies because of interaction enhancement.

\begin{figure}
\includegraphics[width=8cm,clip=]{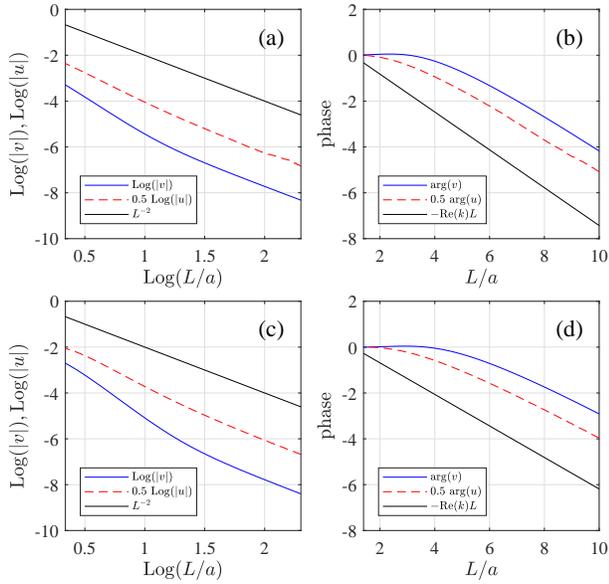}
\caption{Dependence of the matrix elements $v_n$ and $u_n$ on the
stance between disks $L$. (a) and (b) corresponds to Fig.
\ref{fig3} (d) and (c) and (d) does to Fig. \ref{fig5} (b).}
\label{fig4}
\end{figure}

In order to quantitatively evaluate this interaction we start
consideration with an isolated disk for which the matrix of
derivatives in Eq. (\ref{ME}) becomes diagonal with the complex
eigenfrequencies $k_n$ in the eigenbasis presented in Fig.
\ref{fig2}. It is reasonable to consider that for enough
separation of two disks the matrix is still diagonal with pairs of
degenerate $k_n$ shown in Fig. \ref{fig3} by blue closed circles.
Rigorously speaking for the large distance between the disks $L
\gg a/\gamma_n$ the interaction via the resonant modes  can grow
exponentially \cite{Lalanne2018}. In view of that we restrict the
distance $L < a/\gamma_n$. As the distance between the disks is
reduced the interaction between the disks via the resonant modes
splits the degenerate resonant modes $k_n$ giving rise to the
avoided crossing. Assume also that a value of splitting much less
than the distance between the different $k_n$. These assumptions
are justified numerically as shown in insets of Fig. \ref{fig3},
however, for only definite domains of the frequency $k$ around the
resonances of the isolated disk where spiral behavior of the
resonant frequencies takes place. In the framework of these
assumptions we can use two-level approximation for the Hamiltonian
matrix in Eq. (\ref{ME}) for each resonance $k_n$
\cite{Wiersig2006,Lalanne2018,Bogdanov2019}
\begin{equation}\label{H0V}
H_{eff}^{(n)}=H_{eff}^{(0)}+V=\left(\begin{array}{cc} k_na& 0\cr
0 &k_na\end{array}\right)+\left(\begin{array}{cc} u_n& v_n\cr
v_n &u_n\end{array}\right),
\end{equation}
where $v_n$ is responsible for interaction between the disks
 via the resonant modes while
$u_n$ is the result of the  backscattering by the first disk.
Therefore one can expect that $arg(v_n)=\omega_nL/a, ~~
arg(u_n)=2\omega_nL/a$. Fig. \ref{fig4} shows the behavior of the
absolute value and phase both of the matrix elements. The matrix
elements $v_n$ and $u_n$ can be easily found from numerically
calculated resonances shown in Fig. \ref{fig3}
\begin{equation}\label{poles}
k_{a,s}^{(n)}a=k_na+u_n \pm v_n,
\end{equation}
as $v_n=\frac{k_s^{(n)}-k_a^{(n)}}{2},
u_n=\frac{k_s^{(n)}+k_a^{(n)}}{2}-k_n$.
From Fig. \ref{fig4} one can evaluate that the interaction term in
(\ref{H0V})
\begin{equation}\label{vu}
    v_n\sim \frac{e^{ik_nL}}{L^2}, u_n\sim \frac{e^{2ik_nL}}{L^4}.
\end{equation}
     The distance behavior (\ref{vu}) is
observed with good accuracy for all resonances shown in Fig.
\ref{fig3}, however, for only spiral convergence of the
resonances. Numerically calculated behavior of the matrix elements
$v_n$ and $u_n$ for $n=2$ is shown in Fig. \ref{fig4}. In
spiralling around the resonances of the isolated disk the
hybridized resonant eigenmode is given by symmetric and
antisymmetric combinations of the resonant modes of the isolated
disk
\begin{equation}\label{s,a}
    \psi_{s,a}(\overrightarrow{r})=\psi_n(\overrightarrow{r}_{\perp}-\frac{1}{2}L\overrightarrow{z})
    \pm \psi_n(\overrightarrow{r}_{\perp}+\frac{1}{2}L\overrightarrow{z})
\end{equation}
where $\overrightarrow{r}_{\perp}=r,\phi$, $\overrightarrow{z}$ is
unit vector along the z-axis and
$\psi_n(\overrightarrow{r}_{\perp})$ are the corresponding
resonant mode of the isolated disk shown in the insets in Fig.
\ref{fig2}.

At first the resonant frequencies slowly spiral away from the
limiting point given by $k_n$. Respectively the $Q$ factor in Fig.
\ref{fig3} (c) demonstrates oscillating behavior exceeding the $Q$
factor of the isolated disk a few times. With approaching of disks
spiral behavior of the pair of resonances $k_{s,a}^{(n)}$ is
replaced by strong repulsion as shown Fig. \ref{fig3} (a). Fig.
\ref{fig3} (d) shows a remarkable feature caused by avoided
crossing of resonances with different $n$. To be specific there is
avoided crossing of symmetric resonances $k_{s}^{(2)}$ and
$k_{s}^{(5)}$ according to enumeration in Fig. \ref{fig2}. Because
of the same symmetry of resonances relative to $z\rightarrow -z$
these resonances undergo typical avoided crossing with a
considerable decrease of the imaginary part of the resonant
frequency and correspondingly enhancement of the $Q$ factor by one
order in magnitude. Respectively the two-mode approximation
(\ref{H0V}) breaks down.

\begin{figure}
\includegraphics[width=9cm,clip=]{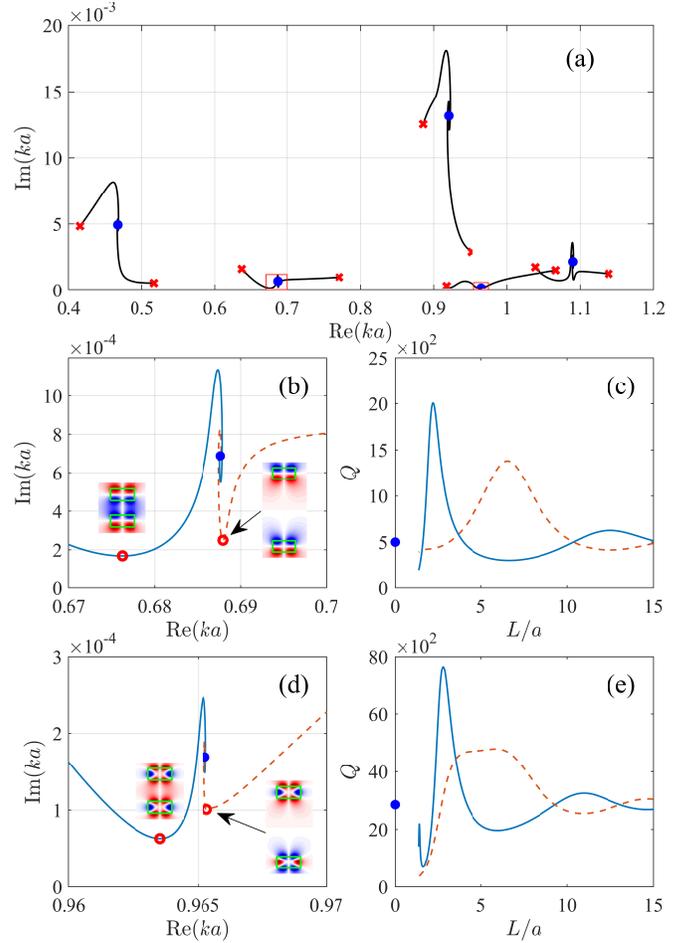}
\caption{(a) Behavior of resonant eigenfrequencies under variation
of the distance between the disks $L$ with $\epsilon=40$ (ceramics
in THz range) and the aspect ratio $a/h=0.7$. (b) and (d) zoomed
areas highlighted in (a) with symmetric (solid lines) and
antisymmetric (dash lines) hybridization (\ref{s,a}) of resonant
modes of the isolated disk. (c) and (e) show behavior of the Q
factor vs the distance for corresponding insets at the left.
Closed circles mark the eigenfrequencies of isolated disks and
respectively the $Q$ factors while crosses mark the limiting case
$L=1$ when two disks stick together.} \label{fig5}
\end{figure}

It is interesting to trace the behavior of resonances and $Q$
factors for the aspect ratio $a/h\approx 0.7$ and $\epsilon=40$
for which the isolated disk shows the maximal $Q$ factor
\cite{Rybin2017,Koshelev2018}. The results are presented in Fig.
\ref{fig5}.
One can see that with decrease of the distance between the disks
we have the same spiralling  behavior of the hybridized resonances
around the resonances of the isolated disks which is terminated by
strong repulsion of the symmetric and antisymmetric resonances for
$L\rightarrow h$. However, we have no pronounced effect of the
avoiding crossing of hybridized resonances with different $n$ and
respectively have no enhancement of the $Q$ factor by one order as
it was achieved for the aspect ratio $a/h=1$ (see Fig. \ref{fig4}
(e)).

\begin{figure}
\includegraphics[width=9cm,clip=]{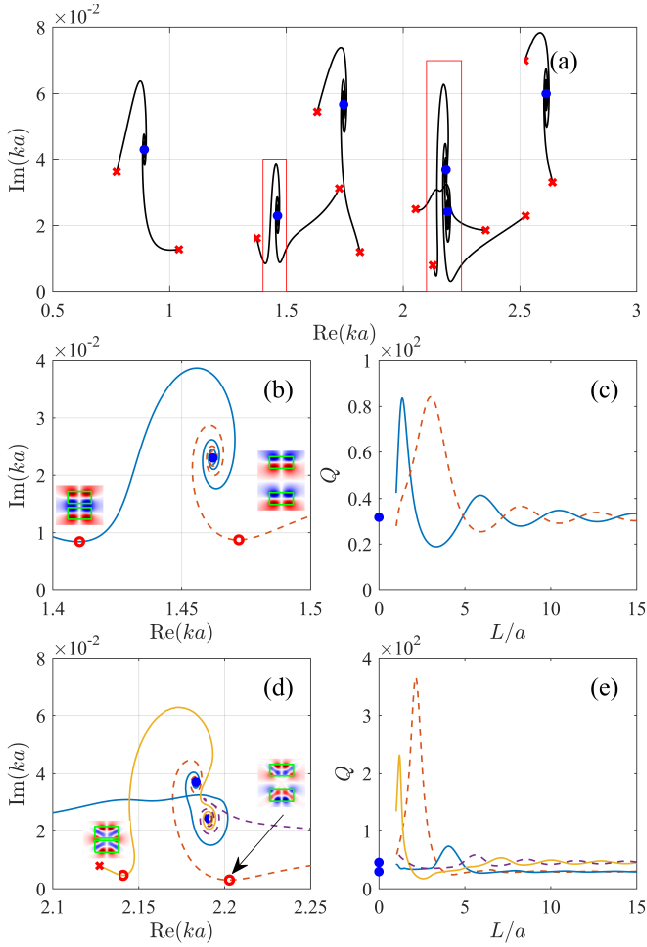}
\caption{(a) Behavior of resonant eigenfrequencies under variation
of the distance between the disks $L$ with $\epsilon=12$ (silica in optical range) and
aspect ratio $a/h=1$. (b) and (d) zoomed
areas highlighted in (a) with symmetric (solid lines) and antisymmetric (dash lines)
hybridization (\ref{s,a}) of resonant modes of the isolated disk.
(c) and (e) show behavior of the $Q$ factor vs the distance for
corresponding insets at the left. Closed circles mark the eigenfrequencies of isolated disks and
respectively the $Q$ factors while crosses mark the limiting case $L=1$ when two disks stick together.}
\label{fig6}
\end{figure}

Till now we considered the permittivity $\epsilon=40$ and $a=1cm$
(ceramic disks) that enters the resonant frequencies into the THz
range. Finally, we consider $\epsilon=12$ (silica disks) and
$a=h=1\mu$ with the resonant frequencies in the optical range.
Results of computations are presented  in Fig. \ref{fig6} which
show that there is no qualitative difference between the ceramic
disks with $\epsilon=40$ and silica disks with $\epsilon=12$.
Similar to Fig. \ref{fig3} and Fig. \ref{fig5} we observe spiral
behavior of the resonant frequencies around the resonances of the
isolated disk
 for the enough distance between the disks. But what is more remarkable also we observe the
 avoided crossing of the resonances with different $n$ as shown in Fig. \ref{fig6} (d) with corresponding
 strong enhancement of the $Q$ factor by one order in magnitude (Fig. \ref{fig6} (e)).

\section{Conclusions}

The recept to enhance the $Q$ factor by means of the avoided
crossing of resonances is well known. Friedrich and Wintgen
\cite{Friedrich1985} were the first who investigated the
quantitative influence of the interference of resonances on their
positions and widths.  Moreover, in the framework of two-level
effective Hamiltonian they found out that one of the widths can
turn to zero to identify the BIC. A single isolated dielectric
particle of finite dimensions can not trap light \cite{Colton}
because of the infinite number of radiation continua or
diffraction channels \cite{PRA96}. However, for sufficiently large
refractive index the particle shows distinctive Mie resonances
with the Q-factors which can be substantially enhanced owing to
the avoided crossing of the resonances under variation of aspect
ratio of the disk \cite{Rybin17,Bogdanov2019}. Technologically it
might be challengeable to vary the size of the disk in the optical
range. In the present paper we propose to vary the distance
between two coaxial disks that is preferable from the experimental
viewpoint. Continuous variation of the distance gives rise to an
avoided crossing of the Mie resonances due to interaction between
the disks through radiating resonant modes.

For the enough distance we can consider that two disks have
degenerate resonances $k_n$. However with drawing closer of the
disks the disks begin weakly interact with each other via the
leaking resonant modes that lifts a degeneracy of the resonances
according to Eq. (\ref{poles}) and respective symmetric and
antisymmetric hybridizations of resonant modes $k_{s,a}^{(n)}$
according to Eq. (\ref{s,a}). As the result we observe "soft"
avoided crossing of the resonances around the points $k_n$. The
further decrease of the distance between the disks enhances the
interaction and respectively gives rise to strong repulsion of
$k_s^{(n)}$ and $k_a^{(n)}$. However what is the most remarkable
there are events of the avoided crossing of the resonances with
different $n$. Respectively one can observe strong enhancement of
the $Q$ factor around one order in magnitude.

Although in the present paper we considered only the dielectric
disks, it is clear that the phenomenon of the avoided crossing and
respective enhancement of the $Q$ factor would occur with particles
of arbitrary shape when the distance between them  is varied. The
case of two coaxial disks simplifies computations because the
solutions with different angular momentum $m$ are independent. In
the present paper we have presented only the case $m=0$ because of
a possibility to consider separately E and H polarizations.

\begin{acknowledgments}
We are grateful to A.A. Bogdanov, E.N. Bulgakov and D.N. Maksimov
for numerous and fruitful discussions.
This  work  was  supported  by RFBR grant 19-02-00055.
\end{acknowledgments}

\bibliography{sadreev}

\begin{thebibliography}{30}%
\makeatletter
\providecommand \@ifxundefined [1]{%
 \@ifx{#1\undefined}
}%
\providecommand \@ifnum [1]{%
 \ifnum #1\expandafter \@firstoftwo
 \else \expandafter \@secondoftwo
 \fi
}%
\providecommand \@ifx [1]{%
 \ifx #1\expandafter \@firstoftwo
 \else \expandafter \@secondoftwo
 \fi
}%
\providecommand \natexlab [1]{#1}%
\providecommand \enquote  [1]{``#1''}%
\providecommand \bibnamefont  [1]{#1}%
\providecommand \bibfnamefont [1]{#1}%
\providecommand \citenamefont [1]{#1}%
\providecommand \href@noop [0]{\@secondoftwo}%
\providecommand \href [0]{\begingroup \@sanitize@url \@href}%
\providecommand \@href[1]{\@@startlink{#1}\@@href}%
\providecommand \@@href[1]{\endgroup#1\@@endlink}%
\providecommand \@sanitize@url [0]{\catcode `\\12\catcode `\$12\catcode
  `\&12\catcode `\#12\catcode `\^12\catcode `\_12\catcode `\%12\relax}%
\providecommand \@@startlink[1]{}%
\providecommand \@@endlink[0]{}%
\providecommand \url  [0]{\begingroup\@sanitize@url \@url }%
\providecommand \@url [1]{\endgroup\@href {#1}{\urlprefix }}%
\providecommand \urlprefix  [0]{URL }%
\providecommand \Eprint [0]{\href }%
\providecommand \doibase [0]{http://dx.doi.org/}%
\providecommand \selectlanguage [0]{\@gobble}%
\providecommand \bibinfo  [0]{\@secondoftwo}%
\providecommand \bibfield  [0]{\@secondoftwo}%
\providecommand \translation [1]{[#1]}%
\providecommand \BibitemOpen [0]{}%
\providecommand \bibitemStop [0]{}%
\providecommand \bibitemNoStop [0]{.\EOS\space}%
\providecommand \EOS [0]{\spacefactor3000\relax}%
\providecommand \BibitemShut  [1]{\csname bibitem#1\endcsname}%
\let\auto@bib@innerbib\@empty
\bibitem [{\citenamefont {Braginsky}, \citenamefont {Gorodetsky},\ and\
  \citenamefont {Ilchenko}(1989)}]{Braginsky1989}%
  \BibitemOpen
  \bibfield  {author} {\bibinfo {author} {\bibfnamefont {V.}~\bibnamefont
  {Braginsky}}, \bibinfo {author} {\bibfnamefont {M.}~\bibnamefont
  {Gorodetsky}}, \ and\ \bibinfo {author} {\bibfnamefont {V.}~\bibnamefont
  {Ilchenko}},\ }\bibfield  {title} {\enquote {\bibinfo {title} {Quality-factor
  and nonlinear properties of optical whispering-gallery modes},}\ }\href
  {\doibase 10.1016/0375-9601(89)90912-2} {\bibfield  {journal} {\bibinfo
  {journal} {Phys. Let. A}\ }\textbf {\bibinfo {volume} {137}},\ \bibinfo
  {pages} {393} (\bibinfo {year} {1989})}\BibitemShut {NoStop}%
\bibitem [{\citenamefont {Cao}\ and\ \citenamefont {Wiersig}(2015)}]{Cao2015}%
  \BibitemOpen
  \bibfield  {author} {\bibinfo {author} {\bibfnamefont {H.}~\bibnamefont
  {Cao}}\ and\ \bibinfo {author} {\bibfnamefont {J.}~\bibnamefont {Wiersig}},\
  }\bibfield  {title} {\enquote {\bibinfo {title} {Dielectric microcavities:
  Model systems for wave chaos and non-hermitian physics},}\ }\href {\doibase
  10.1103/revmodphys.87.61} {\bibfield  {journal} {\bibinfo  {journal} {Rev.
  Mod. Phys.}\ }\textbf {\bibinfo {volume} {87}},\ \bibinfo {pages} {61}
  (\bibinfo {year} {2015})}\BibitemShut {NoStop}%
\bibitem [{\citenamefont {Kuznetsov}\ \emph {et~al.}(2016)\citenamefont
  {Kuznetsov}, \citenamefont {Miroshnichenko}, \citenamefont {Brongersma},
  \citenamefont {Kivshar},\ and\ \citenamefont {Luk'yanchuk}}]{Kuznetsov2016}%
  \BibitemOpen
  \bibfield  {author} {\bibinfo {author} {\bibfnamefont {A.}~\bibnamefont
  {Kuznetsov}}, \bibinfo {author} {\bibfnamefont {A.}~\bibnamefont
  {Miroshnichenko}}, \bibinfo {author} {\bibfnamefont {M.}~\bibnamefont
  {Brongersma}}, \bibinfo {author} {\bibfnamefont {Y.}~\bibnamefont {Kivshar}},
  \ and\ \bibinfo {author} {\bibfnamefont {B.}~\bibnamefont {Luk'yanchuk}},\
  }\bibfield  {title} {\enquote {\bibinfo {title} {Optically resonant
  dielectric nanostructures},}\ }\href {\doibase 10.1126/science.aag2472}
  {\bibfield  {journal} {\bibinfo  {journal} {Science}\ }\textbf {\bibinfo
  {volume} {354}},\ \bibinfo {pages} {2472} (\bibinfo {year}
  {2016})}\BibitemShut {NoStop}%
\bibitem [{\citenamefont {Koshelev}, \citenamefont {Bogdanov},\ and\
  \citenamefont {Kivshar}(2018)}]{Koshelev2018}%
  \BibitemOpen
  \bibfield  {author} {\bibinfo {author} {\bibfnamefont {K.}~\bibnamefont
  {Koshelev}}, \bibinfo {author} {\bibfnamefont {A.}~\bibnamefont {Bogdanov}},
  \ and\ \bibinfo {author} {\bibfnamefont {Y.}~\bibnamefont {Kivshar}},\
  }\bibfield  {title} {\enquote {\bibinfo {title} {Meta-optics and bound states
  in the continuum},}\ }\href@noop {} {\bibfield  {journal} {\bibinfo
  {journal} {Science Bulletin}\ }\textbf {\bibinfo {volume} {17}},\ \bibinfo
  {pages} {065601} (\bibinfo {year} {2018})}\BibitemShut {NoStop}%
\bibitem [{\citenamefont {Friedrich}\ and\ \citenamefont
  {Wintgen}(1985)}]{Friedrich1985}%
  \BibitemOpen
  \bibfield  {author} {\bibinfo {author} {\bibfnamefont {H.}~\bibnamefont
  {Friedrich}}\ and\ \bibinfo {author} {\bibfnamefont {D.}~\bibnamefont
  {Wintgen}},\ }\bibfield  {title} {\enquote {\bibinfo {title} {Interfering
  resonances and bound states in the continuum},}\ }\href {\doibase
  10.1103/physreva.32.3231} {\bibfield  {journal} {\bibinfo  {journal} {Phys.
  Rev. A}\ }\textbf {\bibinfo {volume} {32}},\ \bibinfo {pages} {3231}
  (\bibinfo {year} {1985})}\BibitemShut {NoStop}%
\bibitem [{\citenamefont {Sadreev}, \citenamefont {Bulgakov},\ and\
  \citenamefont {Rotter}(2006)}]{SBR}%
  \BibitemOpen
  \bibfield  {author} {\bibinfo {author} {\bibfnamefont {A.}~\bibnamefont
  {Sadreev}}, \bibinfo {author} {\bibfnamefont {E.}~\bibnamefont {Bulgakov}}, \
  and\ \bibinfo {author} {\bibfnamefont {I.}~\bibnamefont {Rotter}},\
  }\bibfield  {title} {\enquote {\bibinfo {title} {Bound states in the
  continuum in open quantum billiards with a variable shape},}\ }\href@noop {}
  {\bibfield  {journal} {\bibinfo  {journal} {Phys. Rev. B}\ }\textbf {\bibinfo
  {volume} {73}},\ \bibinfo {pages} {235342} (\bibinfo {year}
  {2006})}\BibitemShut {NoStop}%
\bibitem [{\citenamefont {Bulgakov}\ and\ \citenamefont
  {Sadreev}(2008)}]{BS2008}%
  \BibitemOpen
  \bibfield  {author} {\bibinfo {author} {\bibfnamefont {E.}~\bibnamefont
  {Bulgakov}}\ and\ \bibinfo {author} {\bibfnamefont {A.}~\bibnamefont
  {Sadreev}},\ }\bibfield  {title} {\enquote {\bibinfo {title} {Bound states in
  the continuum in photonic waveguides inspired by defects},}\ }\href {\doibase
  10.1103/PhysRevB.78.075105} {\bibfield  {journal} {\bibinfo  {journal} {Phys.
  Rev. B}\ }\textbf {\bibinfo {volume} {78}},\ \bibinfo {pages} {075105}
  (\bibinfo {year} {2008})}\BibitemShut {NoStop}%
\bibitem [{\citenamefont {Shipman}\ and\ \citenamefont
  {Venakides}(2005)}]{Shipman2005}%
  \BibitemOpen
  \bibfield  {author} {\bibinfo {author} {\bibfnamefont {S.~P.}\ \bibnamefont
  {Shipman}}\ and\ \bibinfo {author} {\bibfnamefont {S.}~\bibnamefont
  {Venakides}},\ }\bibfield  {title} {\enquote {\bibinfo {title} {Resonant
  transmission near nonrobust periodic slab modes},}\ }\href {\doibase
  10.1103/physreve.71.026611} {\bibfield  {journal} {\bibinfo  {journal} {Phys.
  Rev. E}\ }\textbf {\bibinfo {volume} {71}},\ \bibinfo {pages} {026611}
  (\bibinfo {year} {2005})}\BibitemShut {NoStop}%
\bibitem [{\citenamefont {Marinica}, \citenamefont {Borisov},\ and\
  \citenamefont {Shabanov}(2008)}]{Marinica}%
  \BibitemOpen
  \bibfield  {author} {\bibinfo {author} {\bibfnamefont {D.~C.}\ \bibnamefont
  {Marinica}}, \bibinfo {author} {\bibfnamefont {A.~G.}\ \bibnamefont
  {Borisov}}, \ and\ \bibinfo {author} {\bibfnamefont {S.~V.}\ \bibnamefont
  {Shabanov}},\ }\bibfield  {title} {\enquote {\bibinfo {title} {Bound states
  in the continuum in photonics},}\ }\href@noop {} {\bibfield  {journal}
  {\bibinfo  {journal} {Phys. Rev. Lett.}\ }\textbf {\bibinfo {volume} {100}},\
  \bibinfo {pages} {183902} (\bibinfo {year} {2008})}\BibitemShut {NoStop}%
\bibitem [{\citenamefont {{Chia Wei Hsu}}\ \emph {et~al.}(2013)\citenamefont
  {{Chia Wei Hsu}}, \citenamefont {{Bo Zhen}}, \citenamefont {{Jeongwon Lee}},
  \citenamefont {Johnson}, \citenamefont {Joannopoulos},\ and\ \citenamefont
  {Solja{\v{c}}i{\'c}}}]{Hsu2013}%
  \BibitemOpen
  \bibfield  {author} {\bibinfo {author} {\bibnamefont {{Chia Wei Hsu}}},
  \bibinfo {author} {\bibnamefont {{Bo Zhen}}}, \bibinfo {author} {\bibnamefont
  {{Jeongwon Lee}}}, \bibinfo {author} {\bibfnamefont {S.~G.}\ \bibnamefont
  {Johnson}}, \bibinfo {author} {\bibfnamefont {J.~D.}\ \bibnamefont
  {Joannopoulos}}, \ and\ \bibinfo {author} {\bibfnamefont {M.}~\bibnamefont
  {Solja{\v{c}}i{\'c}}},\ }\bibfield  {title} {\enquote {\bibinfo {title}
  {Observation of trapped light within the radiation continuum},}\ }\href
  {\doibase 10.1038/nature12289} {\bibfield  {journal} {\bibinfo  {journal}
  {Nature}\ }\textbf {\bibinfo {volume} {499}},\ \bibinfo {pages} {188}
  (\bibinfo {year} {2013})}\BibitemShut {NoStop}%
\bibitem [{\citenamefont {Bulgakov}\ and\ \citenamefont
  {Sadreev}(2014)}]{PRA2014}%
  \BibitemOpen
  \bibfield  {author} {\bibinfo {author} {\bibfnamefont {E.~N.}\ \bibnamefont
  {Bulgakov}}\ and\ \bibinfo {author} {\bibfnamefont {A.~F.}\ \bibnamefont
  {Sadreev}},\ }\bibfield  {title} {\enquote {\bibinfo {title} {Bloch bound
  states in the radiation continuum in a periodic array of dielectric rods},}\
  }\href {\doibase 10.1103/physreva.90.053801} {\bibfield  {journal} {\bibinfo
  {journal} {Phys. Rev. A}\ }\textbf {\bibinfo {volume} {90}},\ \bibinfo
  {pages} {053801} (\bibinfo {year} {2014})}\BibitemShut {NoStop}%
\bibitem [{\citenamefont {Bulgakov}\ and\ \citenamefont
  {Sadreev}(2017{\natexlab{a}})}]{PRA96}%
  \BibitemOpen
  \bibfield  {author} {\bibinfo {author} {\bibfnamefont {E.}~\bibnamefont
  {Bulgakov}}\ and\ \bibinfo {author} {\bibfnamefont {A.~F.}\ \bibnamefont
  {Sadreev}},\ }\bibfield  {title} {\enquote {\bibinfo {title} {Bound states in
  the continuum with high orbital angular momentum in a dielectric rod with
  periodically modulated permittivity},}\ }\href@noop {} {\bibfield  {journal}
  {\bibinfo  {journal} {Phys. Rev. A}\ }\textbf {\bibinfo {volume} {96}},\
  \bibinfo {pages} {013841} (\bibinfo {year} {2017}{\natexlab{a}})}\BibitemShut
  {NoStop}%
\bibitem [{\citenamefont {Colton}\ and\ \citenamefont
  {R.Kress}(1998)}]{Colton}%
  \BibitemOpen
  \bibfield  {author} {\bibinfo {author} {\bibfnamefont {D.}~\bibnamefont
  {Colton}}\ and\ \bibinfo {author} {\bibnamefont {R.Kress}},\ }\href@noop {}
  {\emph {\bibinfo {title} {Inverse Acoustic and Electromagnetic Scattering
  Theory}}},\ \bibinfo {edition} {2nd}\ ed.\ (\bibinfo  {publisher}
  {Springer,Berlin},\ \bibinfo {year} {1998})\BibitemShut {NoStop}%
\bibitem [{\citenamefont {Silveirinha}(2014)}]{Silveirinha14}%
  \BibitemOpen
  \bibfield  {author} {\bibinfo {author} {\bibfnamefont {M.~G.}\ \bibnamefont
  {Silveirinha}},\ }\bibfield  {title} {\enquote {\bibinfo {title} {Trapping
  light in open plasmonic nanostructures},}\ }\href {\doibase
  10.1103/physreva.89.023813} {\bibfield  {journal} {\bibinfo  {journal} {Phys.
  Rev. A}\ }\textbf {\bibinfo {volume} {89}},\ \bibinfo {pages} {023813}
  (\bibinfo {year} {2014})}\BibitemShut {NoStop}%
\bibitem [{\citenamefont {Balyzin}\ \emph {et~al.}(2018)\citenamefont
  {Balyzin}, \citenamefont {Sadrieva}, \citenamefont {Belyakov}, \citenamefont
  {Kapitanova}, \citenamefont {Sadreev},\ and\ \citenamefont
  {Bogdanov}}]{Balyzin2018}%
  \BibitemOpen
  \bibfield  {author} {\bibinfo {author} {\bibfnamefont {M.}~\bibnamefont
  {Balyzin}}, \bibinfo {author} {\bibfnamefont {Z.}~\bibnamefont {Sadrieva}},
  \bibinfo {author} {\bibfnamefont {M.}~\bibnamefont {Belyakov}}, \bibinfo
  {author} {\bibfnamefont {P.}~\bibnamefont {Kapitanova}}, \bibinfo {author}
  {\bibfnamefont {A.}~\bibnamefont {Sadreev}}, \ and\ \bibinfo {author}
  {\bibfnamefont {A.}~\bibnamefont {Bogdanov}},\ }\bibfield  {title} {\enquote
  {\bibinfo {title} {Bound state in the continuum in 1d chain of dielectric
  disks: theory and experiment},}\ }\href {\doibase
  10.1088/1742-6596/1092/1/012012} {\bibfield  {journal} {\bibinfo  {journal}
  {J. Phys.: Conf. Ser.}\ }\textbf {\bibinfo {volume} {1092}},\ \bibinfo
  {pages} {012012} (\bibinfo {year} {2018})}\BibitemShut {NoStop}%
\bibitem [{\citenamefont {Rybin}\ \emph {et~al.}(2017)\citenamefont {Rybin},
  \citenamefont {Koshelev}, \citenamefont {Sadrieva}, \citenamefont {Samusev},
  \citenamefont {Bogdanov}, \citenamefont {Limonov},\ and\ \citenamefont
  {Kivshar}}]{Rybin2017}%
  \BibitemOpen
  \bibfield  {author} {\bibinfo {author} {\bibfnamefont {M.~V.}\ \bibnamefont
  {Rybin}}, \bibinfo {author} {\bibfnamefont {K.~L.}\ \bibnamefont {Koshelev}},
  \bibinfo {author} {\bibfnamefont {Z.~F.}\ \bibnamefont {Sadrieva}}, \bibinfo
  {author} {\bibfnamefont {K.~B.}\ \bibnamefont {Samusev}}, \bibinfo {author}
  {\bibfnamefont {A.~A.}\ \bibnamefont {Bogdanov}}, \bibinfo {author}
  {\bibfnamefont {M.~F.}\ \bibnamefont {Limonov}}, \ and\ \bibinfo {author}
  {\bibfnamefont {Y.~S.}\ \bibnamefont {Kivshar}},\ }\bibfield  {title}
  {\enquote {\bibinfo {title} {High-{Q} {S}upercavity {M}odes in
  {S}ubwavelength {D}ielectric {R}esonators},}\ }\href {\doibase
  10.1103/physrevlett.119.243901} {\bibfield  {journal} {\bibinfo  {journal}
  {Phys. Rev. Lett.}\ }\textbf {\bibinfo {volume} {119}},\ \bibinfo {pages}
  {243901} (\bibinfo {year} {2017})}\BibitemShut {NoStop}%
\bibitem [{\citenamefont {Chen}, \citenamefont {Chen},\ and\ \citenamefont
  {Liu}(2018)}]{WeijinChen2018}%
  \BibitemOpen
  \bibfield  {author} {\bibinfo {author} {\bibfnamefont {W.}~\bibnamefont
  {Chen}}, \bibinfo {author} {\bibfnamefont {Y.}~\bibnamefont {Chen}}, \ and\
  \bibinfo {author} {\bibfnamefont {W.}~\bibnamefont {Liu}},\ }\href@noop {}
  {\enquote {\bibinfo {title} {Subwavelength {H}igh-{Q} {K}erker {S}uper modes
  with {U}nidirectional {R}adiations},}\ } (\bibinfo {year} {2018}),\ \bibinfo
  {note} {arXiv preprint arXiv:1808.05539}\BibitemShut {NoStop}%
\bibitem [{\citenamefont {Bogdanov}\ \emph {et~al.}(2019)\citenamefont
  {Bogdanov}, \citenamefont {Koshelev}, \citenamefont {Kapitanova},
  \citenamefont {Rybin}, \citenamefont {Gladyshev}, \citenamefont {Sadrieva},
  \citenamefont {Samusev}, \citenamefont {Kivshar},\ and\ \citenamefont
  {Limonov}}]{Bogdanov2019}%
  \BibitemOpen
  \bibfield  {author} {\bibinfo {author} {\bibfnamefont {A.}~\bibnamefont
  {Bogdanov}}, \bibinfo {author} {\bibfnamefont {K.}~\bibnamefont {Koshelev}},
  \bibinfo {author} {\bibfnamefont {P.}~\bibnamefont {Kapitanova}}, \bibinfo
  {author} {\bibfnamefont {M.}~\bibnamefont {Rybin}}, \bibinfo {author}
  {\bibfnamefont {S.}~\bibnamefont {Gladyshev}}, \bibinfo {author}
  {\bibfnamefont {Z.}~\bibnamefont {Sadrieva}}, \bibinfo {author}
  {\bibfnamefont {K.}~\bibnamefont {Samusev}}, \bibinfo {author} {\bibfnamefont
  {Y.}~\bibnamefont {Kivshar}}, \ and\ \bibinfo {author} {\bibfnamefont
  {M.~F.}\ \bibnamefont {Limonov}},\ }\bibfield  {title} {\enquote {\bibinfo
  {title} {Bound states in the continuum and fano resonances in the strong mode
  coupling regime},}\ }\href {\doibase 10.1117/1.ap.1.1.016001} {\bibfield
  {journal} {\bibinfo  {journal} {Adv. Photonics}\ }\textbf {\bibinfo {volume}
  {1}},\ \bibinfo {pages} {1} (\bibinfo {year} {2019})}\BibitemShut {NoStop}%
\bibitem [{\citenamefont {Wiersig}(2006)}]{Wiersig2006}%
  \BibitemOpen
  \bibfield  {author} {\bibinfo {author} {\bibfnamefont {J.}~\bibnamefont
  {Wiersig}},\ }\bibfield  {title} {\enquote {\bibinfo {title} {Formation of
  {L}ong-{L}ived, {S}carlike {M}odes near {A}voided {R}esonance {C}rossings in
  {O}ptical {M}icrocavities},}\ }\href@noop {} {\bibfield  {journal} {\bibinfo
  {journal} {Phys. Rev. Lett.}\ }\textbf {\bibinfo {volume} {97}} (\bibinfo
  {year} {2006})}\BibitemShut {NoStop}%
\bibitem [{\citenamefont {Boriskina}(2006)}]{Boriskina2006}%
  \BibitemOpen
  \bibfield  {author} {\bibinfo {author} {\bibfnamefont {S.~V.}\ \bibnamefont
  {Boriskina}},\ }\bibfield  {title} {\enquote {\bibinfo {title} {Theoretical
  prediction of a dramatic q-factor enhancement and degeneracy removal of
  whispering gallery modes in symmetrical photonic molecules},}\ }\href
  {\doibase 10.1364/ol.31.000338} {\bibfield  {journal} {\bibinfo  {journal}
  {Opt. Lett.}\ }\textbf {\bibinfo {volume} {31}},\ \bibinfo {pages} {338}
  (\bibinfo {year} {2006})}\BibitemShut {NoStop}%
\bibitem [{\citenamefont {Boriskina}(2007)}]{Boriskina2007}%
  \BibitemOpen
  \bibfield  {author} {\bibinfo {author} {\bibfnamefont {S.}~\bibnamefont
  {Boriskina}},\ }\bibfield  {title} {\enquote {\bibinfo {title} {Coupling of
  whispering-gallery modes in size-mismatched microdisk photonic molecules},}\
  }\href {\doibase 10.1364/ol.32.001557} {\bibfield  {journal} {\bibinfo
  {journal} {Opt. Lett.}\ }\textbf {\bibinfo {volume} {32}},\ \bibinfo {pages}
  {1557} (\bibinfo {year} {2007})}\BibitemShut {NoStop}%
\bibitem [{\citenamefont {Unterhinninghofen}, \citenamefont {Wiersig},\ and\
  \citenamefont {Hentschel}(2008)}]{Unter2008}%
  \BibitemOpen
  \bibfield  {author} {\bibinfo {author} {\bibfnamefont {J.}~\bibnamefont
  {Unterhinninghofen}}, \bibinfo {author} {\bibfnamefont {J.}~\bibnamefont
  {Wiersig}}, \ and\ \bibinfo {author} {\bibfnamefont {M.}~\bibnamefont
  {Hentschel}},\ }\bibfield  {title} {\enquote {\bibinfo {title} {Goos-hanchen
  shift and localization of optical modes in deformed microcavities},}\
  }\href@noop {} {\bibfield  {journal} {\bibinfo  {journal} {Phys. Rev. E}\
  }\textbf {\bibinfo {volume} {78}} (\bibinfo {year} {2008})}\BibitemShut
  {NoStop}%
\bibitem [{\citenamefont {Benyoucef}\ \emph {et~al.}(2011)\citenamefont
  {Benyoucef}, \citenamefont {Shim}, \citenamefont {Wiersig},\ and\
  \citenamefont {Schmidt}}]{Benyoucef2011}%
  \BibitemOpen
  \bibfield  {author} {\bibinfo {author} {\bibfnamefont {M.}~\bibnamefont
  {Benyoucef}}, \bibinfo {author} {\bibfnamefont {J.-B.}\ \bibnamefont {Shim}},
  \bibinfo {author} {\bibfnamefont {J.}~\bibnamefont {Wiersig}}, \ and\
  \bibinfo {author} {\bibfnamefont {O.~G.}\ \bibnamefont {Schmidt}},\
  }\bibfield  {title} {\enquote {\bibinfo {title} {Quality-factor enhancement
  of supermodes in coupled microdisks},}\ }\href {\doibase
  10.1364/ol.36.001317} {\bibfield  {journal} {\bibinfo  {journal} {Opt.
  Lett.}\ }\textbf {\bibinfo {volume} {36}},\ \bibinfo {pages} {1317} (\bibinfo
  {year} {2011})}\BibitemShut {NoStop}%
\bibitem [{\citenamefont {Bulgakov}\ and\ \citenamefont
  {Sadreev}(2017{\natexlab{b}})}]{Bulgakov2017}%
  \BibitemOpen
  \bibfield  {author} {\bibinfo {author} {\bibfnamefont {E.~N.}\ \bibnamefont
  {Bulgakov}}\ and\ \bibinfo {author} {\bibfnamefont {A.~F.}\ \bibnamefont
  {Sadreev}},\ }\bibfield  {title} {\enquote {\bibinfo {title} {Bound states in
  the continuum with high orbital angular momentum in a dielectric rod with
  periodically modulated permittivity},}\ }\href {\doibase
  10.1103/physreva.96.013841} {\bibfield  {journal} {\bibinfo  {journal} {Phys.
  Rev. A}\ }\textbf {\bibinfo {volume} {96}},\ \bibinfo {pages} {013841}
  (\bibinfo {year} {2017}{\natexlab{b}})}\BibitemShut {NoStop}%
\bibitem [{\citenamefont {Joannopoulos}(1995)}]{Joan}%
  \BibitemOpen
  \bibfield  {author} {\bibinfo {author} {\bibfnamefont {M.~R. D. W. J.~N.}\
  \bibnamefont {Joannopoulos}, \bibfnamefont {J.~D.}},\ }\href@noop {} {\emph
  {\bibinfo {title} {Photonic Crystals: Molding the Flow of Light}}}\ (\bibinfo
   {publisher} {Princeton Univ.Press},\ \bibinfo {address} {Princeton, NJ,},\
  \bibinfo {year} {1995})\BibitemShut {NoStop}%
\bibitem [{\citenamefont {Lalanne}\ \emph {et~al.}(2018)\citenamefont
  {Lalanne}, \citenamefont {Yan}, \citenamefont {Vynck}, \citenamefont
  {Sauvan},\ and\ \citenamefont {Hugonin}}]{Lalanne2018}%
  \BibitemOpen
  \bibfield  {author} {\bibinfo {author} {\bibfnamefont {P.}~\bibnamefont
  {Lalanne}}, \bibinfo {author} {\bibfnamefont {W.}~\bibnamefont {Yan}},
  \bibinfo {author} {\bibfnamefont {K.}~\bibnamefont {Vynck}}, \bibinfo
  {author} {\bibfnamefont {C.}~\bibnamefont {Sauvan}}, \ and\ \bibinfo {author}
  {\bibfnamefont {J.-P.}\ \bibnamefont {Hugonin}},\ }\bibfield  {title}
  {\enquote {\bibinfo {title} {Light interaction with photonic and plasmonic
  resonances},}\ }\href {\doibase 10.1002/lpor.201700113} {\bibfield  {journal}
  {\bibinfo  {journal} {Laser {\&} Photonics Reviews}\ }\textbf {\bibinfo
  {volume} {12}},\ \bibinfo {pages} {1700113} (\bibinfo {year}
  {2018})}\BibitemShut {NoStop}%
\bibitem [{\citenamefont {More}\ and\ \citenamefont
  {Gerjuoy}(1973)}]{More1973}%
  \BibitemOpen
  \bibfield  {author} {\bibinfo {author} {\bibfnamefont {R.~M.}\ \bibnamefont
  {More}}\ and\ \bibinfo {author} {\bibfnamefont {E.}~\bibnamefont {Gerjuoy}},\
  }\bibfield  {title} {\enquote {\bibinfo {title} {Properties of resonance wave
  functions},}\ }\href {\doibase 10.1103/physreva.7.1288} {\bibfield  {journal}
  {\bibinfo  {journal} {Physical Review A}\ }\textbf {\bibinfo {volume} {7}},\
  \bibinfo {pages} {1288--1303} (\bibinfo {year} {1973})}\BibitemShut {NoStop}%
\bibitem [{\citenamefont {Muljarov}, \citenamefont {Langbein},\ and\
  \citenamefont {Zimmermann}(2010)}]{Muljarov2010}%
  \BibitemOpen
  \bibfield  {author} {\bibinfo {author} {\bibfnamefont {E.~A.}\ \bibnamefont
  {Muljarov}}, \bibinfo {author} {\bibfnamefont {W.}~\bibnamefont {Langbein}},
  \ and\ \bibinfo {author} {\bibfnamefont {R.}~\bibnamefont {Zimmermann}},\
  }\bibfield  {title} {\enquote {\bibinfo {title} {Brillouin-wigner
  perturbation theory in open electromagnetic systems},}\ }\href {\doibase
  10.1209/0295-5075/92/50010} {\bibfield  {journal} {\bibinfo  {journal}
  {Europhysics Letters}\ }\textbf {\bibinfo {volume} {92}},\ \bibinfo {pages}
  {50010} (\bibinfo {year} {2010})}\BibitemShut {NoStop}%
\bibitem [{\citenamefont {Snyder}\ and\ \citenamefont
  {Love}(1984)}]{Snyder1984}%
  \BibitemOpen
  \bibfield  {author} {\bibinfo {author} {\bibfnamefont {A.~W.}\ \bibnamefont
  {Snyder}}\ and\ \bibinfo {author} {\bibfnamefont {J.~D.}\ \bibnamefont
  {Love}},\ }\href {\doibase 10.1007/978-1-4613-2813-1} {\emph {\bibinfo
  {title} {Optical Waveguide Theory}}}\ (\bibinfo  {publisher} {Springer
  {US}},\ \bibinfo {year} {1984})\BibitemShut {NoStop}%
\bibitem [{\citenamefont {Rybin}\ and\ \citenamefont
  {Kivshar}(2017)}]{Rybin17}%
  \BibitemOpen
  \bibfield  {author} {\bibinfo {author} {\bibfnamefont {M.}~\bibnamefont
  {Rybin}}\ and\ \bibinfo {author} {\bibfnamefont {Y.}~\bibnamefont
  {Kivshar}},\ }\bibfield  {title} {\enquote {\bibinfo {title} {Optical
  physics: Supercavity lasing},}\ }\href {\doibase 10.1038/541164a} {\bibfield
  {journal} {\bibinfo  {journal} {Nature}\ }\textbf {\bibinfo {volume} {541}},\
  \bibinfo {pages} {164} (\bibinfo {year} {2017})}\BibitemShut {NoStop}%
\end{thebibliography}%

\end{document}